\documentstyle[11pt,newpasp,twoside]{article}
\def\edcomment#1{\iffalse\marginpar{\raggedright\sl#1\/}\else\relax\fi}
\reversemarginpar

\begin{document}
\title{From Young to Old: Spectral Models for Star Cluster Systems}
 \author{Uta Fritze - v. Alvensleben, Peter Anders}
\affil{Universit\"atssternwarte G\"ottingen, Geismarlandstr. 11, 37083 G\"ottingen, Germany}
\author{Richard de Grijs}
\affil{Department of Physics \& Astronomy, University of Sheffield, Hicks Building, Hounsfield Road, Sheffield S3 7RH, UK}

\begin{abstract}
Evolutionary synthesis models for star clusters of various metallicities
including gaseous emission during the lifetime of the ionising stars are
used to model star cluster systems comprising two populations: an old
metal-poor globular clusters (GC) population similar to that of the 
Milky Way halo and a second GC population of arbitrary metallicity. 
We investigate the time evolution of color distributions and 
luminosity functions for the two GC populations
in comparison with observations on E/S0 galaxies. We show that multi-passband data for GC populations give clues to the 
relative ages and metallicities of blue and red subpopulations and help constrain formation
scenarios for their parent galaxies. 
\end{abstract}

\section{Motivation}
GCs are among the oldest objects known, their ages being used to constrain the age of the Universe and the Hubble constant. HST detections of rich systems of bright, blue, young and compact star clusters in numerous interacting and starburst galaxies came as a surprise raising the question if at all, and eventually how many of those young compact clusters could be progenitors of old GCs. This motivated us to calculate a new set of evolutionary synthesis models for the spectral and photometric evolution of star clusters from the very youngest to very old ages and extend them to star cluster systems -- to be complemented by dynamical models for evolution, survival and destruction of clusters in galactic potentials. 

\section{Modelling Star Clusters and Star Cluster Systems}
Our evolutionary synthesis models using Padova stellar isochrones {\bf including the thermal-pulsing AGB phase}, model atmosphere spectra, and gaseous emission ({\bf lines and continuum}) for star clusters of  different metallicities provide the time evolution of spectra, luminosities (U, ..., K), M/L-ratios and colors in many filter systems. Including the TP-AGB phase is important for age-dating clusters on the basis of ${\rm V-I}$ (Schulz et al. 2002). Gaseous emission gives important contributions to broad band fluxes (up to 65 \%) at young ages ($\leq 6-20$ Myr, depending on metallicity) (Anders \& Fritze - v. A. 2003). Theoretical calibrations colors $\leftrightarrow$ [Fe/H] agree well with empirical calibrations for $\sim 12$ Gyr old Milky Way GCs with ${\rm [Fe/H] \leq -0.5}$, get significantly non-linear beyond ${\rm [Fe/H] = -0.5}$ and change completely for clusters younger than 12 Gyr. \\
The age-metallicity-extinction degeneracy -- worst for optical colors with optical-NIR colors better splitting up in metallicity and UV-optical colors better splitting up in extinction -- requires a multi-wavelength analysis to disentangle individual star cluster properties. An ASTROVIRTEL project (PI R. de Grijs, CoIs UFvA, G. Gilmore) provides us with this kind of data. They are compared to a grid of model Spectral Energy Distributions for 5 metallicities ${\rm -1.7 \leq [Fe/H] \leq}$ ${\rm ~ +0.4}$, ages from 4 Myr through 14 Gyr, and extinction values ${\rm 0 \leq E(B-V) \leq 1}$ by means of a dedicated analysis tool (Anders et al. 2003a) to yield individual cluster ages, metallicities, E(B$-$V), masses, and their $1~\sigma$ uncertainties (e.g. Anders et al. 2003b). Extensive tests with artificial clusters and varying numbers of filters show that 1) {\sl a priori} assumptions can be very misleading and 2) a long wavelength baseline and good photometric accuracy are essential, and 3) there are good and bad passband combinations: the U-band is important for ages, extinctions and metallicities of {\bf young} clusters and the NIR for metallicities of {\bf old} clusters (cf. Anders et al. {\sl this volume}). 
Under the simplifying assumptions that a secondary GC population has comparable richness and intrinsic widths of its color distributions (={\bf CD}s) and luminosity functions (={\bf LF}s) to those of the uniform blue peak GC population in E/S0 galaxies or the Milky Way halo we investigate for which combinations of ages and metallicities CDs and LFs show detectable bimodality and look similar to what is seen in E/S0 galaxies (Fritze - v. Alvensleben, {\sl submitted}). We find that 1) it is not clear {\sl a priori} if the red peak GCs are older or more metal-rich than the blue peak clusters, and 2) many combinations (age/metallicity), ranging from (age $=$ 13 Gyr, [Fe/H]$=-0.4$) through (age $=$ 2.5 Gyr, [Fe/H]$=+0.4$), can explain the peak ${\rm \langle V-I \rangle_{red} = 1.2}$ observed e.g. in NGC 4472, NGC 4486, NGC 4649 (Larsen et al. 2001) with, however, significantly different ${\rm \langle V-K \rangle=3.1}$ for the former and ${\rm \langle V-K \rangle=3.6}$ for the latter. Hence, K-band observations will tell the difference and giving decisive clues to the galaxies' (violent) formation histories. 
\acknowledgements
We gratefully acknowledge support by ASTROVIRTEL, funded by the European
Commission under HPRI-CT-1999-00081. \\
UFvA \& PA gratefully acknowledge partial travel support from IAU and DFG. 

\end{document}